\documentclass[a4paper,7pt]{article}
\usepackage{authblk}
\usepackage[dvips]{graphicx}
\graphicspath{{noiseimages/}}
\usepackage{xcolor}
\usepackage{amsmath}
\usepackage{amssymb}
\usepackage{a4wide}
\usepackage[english]{babel}
\usepackage{amsmath}
\usepackage{hyperref}
\usepackage[toc,page]{appendix}
\definecolor{urlcolor}{HTML}{790000}
\newcommand{\bra}[1]{\ensuremath{\langle#1|}}
\newcommand{\ket}[1]{\ensuremath{|#1\rangle}}
\newcommand{\scobka}[2]{\ensuremath{\langle#1 |#2\rangle}}
\newcommand{\x}{{\vec x}}
\newcommand{\p}{{\vec p}}
\newcommand{\y}{{\vec y}}
\newcommand{\eps}{\epsilon}
\renewcommand{\Im}{\operatorname{Im}}
\def\bear{\begin{eqnarray}}
\def\ear{\end{eqnarray}\noindent}

\def\2F1{\phantom{}_2\hspace{-.95pt}F_1}
\author[1,2]{E.T.Akhmedov}
\author[1,2]{K.V.Bazarov}
\author[1,2]{D.V.Diakonov}
\author[3,4]{U.Moschella}
\author[1,2,5]{F.K.Popov}
\author[6]{C.Schubert}
\affil[1]{Institutskii per. 9, Moscow Institute of Physics and Technology, 141700, Dolgoprudny, Russia}
\affil[2]{B. Cheremushkinskaya, 25, Institute for Theoretical and Experimental Physics, 117218, Moscow, Russia}
\affil[3]{INFN, Sez di Milano, Via Celoria 16, 20146, Milano - Italy}
\affil[4]{Universit\`a degli Studi dell'Insubria - Dipartimento DiSAT, Via Valleggio 11 - 22100 Como - Italy}
\affil[5]{Department of Physics, Princeton University, Princeton, NJ 08544}
\affil[6]{Instituto de F\'isica y Matem\'aticas 
Universidad Michoacana de San Nicol\'as de Hidalgo 
Edificio C-3, Apdo. Postal 2-82 
C.P. 58040, Morelia, Michoac\'an, M\'exico} 

\title{\textcolor{black}{Propagators and Gaussian effective actions in various patches of de Sitter space}}

\begin{document}

\numberwithin{equation}{section}

\maketitle

\begin{abstract}

We consider time-ordered (or Feynman) propagators between two different $\alpha-$states of a linear de Sitter Quantum Field in the global de Sitter manifold and in the Poincar\'e patch. We separately examine $\alpha-\beta$, In--In and In--Out  propagators and  find the imaginary contribution to the effective actions. The In--In propagators are real  in both the Poincar\'e patch and in the global de Sitter manifold. On the other side the In--Out propagators at coincident points contain finite imaginary contributions in both patches in even dimensions, but they are not equivalent. In odd dimensions in both patches the imaginary contributions are zero. 
For completeness,  we also consider the Static patch and identify in our construction the  state that is equivalent to the Bunch--Davies one in the Poincar\'e patch.  

\end{abstract}
\newpage

\tableofcontents

\newpage

\section{Introduction}

To explain the point of our study let us consider a real massive scalar field in a curved space. Here and below we denote the mass of the scalar field  $\varphi$ as $m$,  metric as $g_{\mu\nu}$ and the modulus of the determinant of metric as $|g|$.
The effective action is defined as:

\begin{align}
e^{i S_\textsl{eff}}=e^{i \int \mathcal{L}_{\textsl{eff}} \ dx} = \int d [ \varphi] e^{i S[\varphi]} , \  \text{where}  \ \ \ S[\varphi]=\int d^Dx \sqrt{|g|}  \left(\partial_\mu \varphi \partial^\mu \varphi - m^2 \varphi^2\right). \label{action}
\end{align}
It is straightforward to see that

\begin{align*}
\frac{\partial}{\partial m^2} \log \int d [ \varphi] e^{i S[\varphi]} =- i \frac{ \int dx \int d [ \varphi] \varphi(x) \varphi(x) e^{i S[\varphi]}}{\int d [ \varphi] e^{i S[\varphi]}  }=-i  \int dx\ G_F(x,x).
\end{align*}
and this allows to  express the effective Lagrangian via the Feynman (T-ordered) propagator:
\begin{align}
\mathcal{L}_{\textsl{eff}}=\int_{\infty}^{m^2} d\bar m^2  \ G_F(x,x).
\label{GtoL}
\end{align}
Since
 $\scobka{\text{Out}}{\text{In}} = \exp({i\int \mathcal{L}_{\textsl{eff}} \, dx})$, when  
 $\mathcal{L}_{\textsl{eff}}$ is real 
 the transition probability  from the In- to the Out- state is equal to one\footnote{\textcolor{black}{Probably it is worth stressing here the following fact. There is an equality for the amplitudes as follows:}
\begin{eqnarray}
\left\langle {\rm Out} \left| \, T e^{- i \, \int\limits_{-\infty}^{+\infty} H_0(t) \, dt}\right| {\rm In}\right\rangle = \scobka{\text{Out}}{\text{In}}     \nonumber 
\end{eqnarray}
\textcolor{black}{only if $H_0$ is time independent and if $\left| {\rm In}\right\rangle$ is the true ground state of the free Hamiltonian. Otherwise an approximate equality between these two amplitudes holds only for the case of a weak background field. We come back to this point in the main body of the paper.}}. But if the effective Lagrangian has an imaginary part the probability of such a transition is not equal to one:
\begin{align}
\boxed{
\bigg| \scobka{\text{Out}}{\text{In}} \bigg|^2 \ne 1
},
\end{align}
which is usually interpreted as a  signal of particle creation. The Feynman  propagators in de Sitter space having an imaginary part at coincident points is the object of study of this note.

The situation in de Sitter space has certain peculiarities that were pointed out  in \cite{Polyakov:2007mm} and \cite{Akhmedov:2009ta}. It is well known that this space has a maximal isometry group. 
When quantizing fields it is natural to try to respect the isometry if possible. In this case the correlation functions depend only on the scalar invariants. 
However, while in Minkowski space there is a unique Poincar\'e invariant ground state of positive energy, in de Sitter space there is a family of states called the alpha--vacua that respect the isometry at  tree--level \cite{spindel,Mottola:1984ar,Allen:1985ux}.
To calculate the above In--Out amplitude one has to specify which states one wishes to consider. It is possible to calculate the amplitude using the T--ordered In--In (or even alpha--alpha) propagator, or to consider the T--ordered In--Out (or even alpha--beta) propagator. Which one should be chosen? We would like to reconsider this question in this paper.

The paper is organized as follows. In  Section \ref{sec2} we derive analogs of the $\alpha$--modes  {\cite{spindel,Mottola:1984ar,Allen:1985ux}} in the Poincar\'e patch and then consider T-ordered propagators corresponding to the evolution from one $\alpha$--state to another. We then compute the imaginary contributions to the effective actions corresponding to the In--In and In--Out propagators.

\textcolor{black}{Section \ref{paircreation} contains the derivation of the  rate of pair creation in the Poincar\'e patch. In particular we observe that the rate is zero in any odd dimensional de Sitter spacetime; the In--In propagator is real and provides vanishing imaginary contribution to the effective action in any dimension. At the same time the In--Out propagator leads to a non--zero creation rate in even dimension.} We explain which propagator is appropriate to consider in the present circumstances.

{In section \ref{glo} we study the global de Sitter manifold. We show how to relate the $\alpha$--modes in global de Sitter to those previously computed in the Poincar\'e patch, namely, we relate those modes which have equivalent tree--level propagators. We observe that in any odd dimensional de Sitter spacetime the In-- and the Out--modes do coincide; also we show that In--modes in global de Sitter do not provide the same two--point functions as the In--modes in Poincar\'e patch in any dimension. The In--modes in Poincar\'e patch, which are frequently referred to as Bunch--Davies modes, correspond to the so called Euclidean modes in global de Sitter space.}

In  section \ref{sta} we derive once more the well known result that the Bunch--Davies state \cite{Bunch:1978yq} is seen in the Static patch as a thermal equilibrium state {\cite{gibbons,sewell,moschella1,moschella2}}, while the ground state of the free Hamiltonian does not respect the de Sitter isometry.

{In the concluding section we present a heuristic explanation why the pair creation rate should be zero in odd dimensions and explain why there still can be non--trivial stess--energy flux although the rate is vanishing.}

\section{Feynman \texorpdfstring{$\alpha-\beta$}{} propagator in the Poincar\'e patch}
\label{sec2}
\subsection{Free modes in the Poincar\'e patch}\label{sec21}

Let us consider a massive scalar field theory in the   Poincar\'e patch of the $D$-dimensional de Sitter spacetime:
\begin{align*}
ds^2=-dt^2+e^{2t}d{\x}^2,
\end{align*}
where we set the radius of the de Sitter manifold to one. In these  coordinates the de Sitter Klein-Gordon (KG) equation for a real, massive, minimally coupled scalar field  
is written as follows:
\begin{align}
\Bigl[ \partial_t^2+
(D-1)\partial_t
-e^{-2t}\triangle+m^2\Bigr]\varphi=0.
\end{align}
By using the conformal time 
\begin{equation}
{e^{-t}}=\eta \label{ct}
\end{equation}
and separating the variables by defining   $ \varphi(\eta,\vec{x})=\eta^\frac{D-1}{2} \, h(p\eta) \, e^{i\vec{p}\vec{x}}$ one obtains that $h(p\eta)$ must solve the  Bessel differential equation:

\begin{align}
\label{a}
\Bigl[\eta^2\partial_\eta^2+\eta\partial_\eta + (p\eta)^2 + \mu^2\Bigl] h(p \eta)=0,
\end{align}
where
\begin{align*}
 \mu^2=m^2-\frac{(D-1)^2}{4}.
 \end{align*}
Below we restrict our attention to the case $m>\frac{D-1}{2}$ ($\mu $ real); \textcolor{black}{with this restriction modes oscillate at future infinity}. The modes we are going to consider  are therefore of the following form:

\begin{align}
\label{modealpha}
u_{\alpha,\p}({\x},\eta)=   \Bigl(\frac{ \eta}{2\pi}\Bigr)^{\frac{D-1}{2}} e^{i{\p} {\x}}\Bigl[ \alpha_1 H_{i\mu}^{(1)}(p\eta) + \alpha_2 H_{i\mu}^{(2)}(p\eta)\Bigr]
\end{align}
where $H^{(1,2)}_{i\mu}$ denote the first and second type Hankel functions and  $\alpha_1$ and $\alpha_2$ are  complex constants.
The mode expansion of the field operator ${\varphi}$ is then as usual 
\begin{align}
\label{aa}
\varphi_{\alpha}({\x},\eta)=\int {d^{D-1}p} \Bigl(u_{\alpha,\p}({\x},\eta) {a}_p+ u^*_{\alpha,\p}({\x},\eta) {a}^\dagger_p \Bigr).
\end{align}
The relevant Wronskians 
 for the Bessel and  the Hankel functions are given by  
\begin{align}
 W \left\{ J_{\nu}(z),J_{-\nu}(z)\right\}   = - \frac{2\sin(\pi\nu)}{\pi z}, 
 \ \ \ 
 W \left\{ H^{(1)}_{\nu}(z),H^{(2)}_{\nu}(z)\right\}     
= - \frac{4 i }{\pi z}. 
\label{useful}
\end{align}
Taking into account the following relations

\begin{equation} H^{(1)}_{-\nu}(z) = e^{i\pi\nu} H^{(1)}_{\nu}(z) , \ \ \  H^{(2)}_{-\nu}(z) = e^{-i\pi\nu} H^{(2)}_{\nu}(z),   \ \ \  {{H^{(1)}_{\nu}}^*(z)} =  H^{(2)}_{{\nu^*}}({z^*}),  \ \ \ \ {{H^{(2)}_{\nu}}^*(z)} =  H^{(1)}_{{\nu^*}}({z^*}) \label{KK}\end{equation}
one obtains the commutation relations 
\begin{eqnarray}
[\varphi(x,\eta),\pi(y,\eta)] 
=-i \eta^{D-2}\frac{4(|\alpha_2|^2 e^{-\mu\pi}-|\alpha_1|^2e^{\mu\pi})}{\pi}  \delta^{D-1}(\x-\y)
\end{eqnarray}
where $\pi=-\partial_\eta \varphi$ (see Eq. (\ref{ct})).
Canonicity  gives
\begin{align}
\label{comp}
|\alpha_1|^2e^{\mu\pi}-|\alpha_2|^2 e^{-\mu\pi}=\frac{\pi}{4}.
\end{align}
The In-- or Bunch--Davis (BD) modes are proportional to the Hankel function of the first kind  $H_{i\mu}^{(1)}$, i.e. $\alpha_2 = 0$; they behave as \textcolor{black}{pure oscillating exponentials} at past infinity: $H_{i\mu}^{(1)}(p\eta) \sim e^{ip\eta}$ for $p\eta \gg \mu$.
The Out--modes are proportional to the Bessel functions  $J_{i\mu}$ and behave as \textcolor{black}{pure oscillating exponentials} at future infinity: $J_{i\mu}(p\eta) \sim e^{-i\mu \eta}$ for $p\eta \ll \mu$.
While BD--modes do approximately diagonalise the Hamiltonian at past infinity of the Poincar\'e patch, the Out--modes do not  diagonalise the Hamiltonian at any time. None of the modes under consideration diagonalise the Hamiltonian for all times. 

Suppose now to have a second family of canonical modes  of the same form as in Eq. (\ref{modealpha}):
\begin{align}
\label{bb}
u_{\beta,\p}({\x},\eta)=   \Bigl(\frac{ \eta}{2\pi}\Bigr)^{\frac{D-1}{2}} e^{i{\p} {\x}}\Bigl[ \beta_1 H_{i\mu}^{(1)}(p\eta)+\beta_2 H_{i\mu}^{(2)}(p\eta)\Bigr].
\end{align}
There is a Bogoliubov transformation of a particularly simple kind: 
\begin{eqnarray}
&&  u_{\beta,\p}({\x},\eta)=   \gamma\, u_{\alpha,\p}({\x},\eta) + \delta \, u^*_{\alpha,-\p}({\x},\eta), \label{modebeta} 
\end{eqnarray}
where 
\begin{eqnarray}
\gamma  =  \frac 4 \pi (\alpha_1^* \beta_1  e^{\pi\mu}-    \alpha_2^* \beta_2 e^{-\pi\mu}), \ \ \ \delta  =  \frac 4 \pi (\alpha_1 \beta_2  -    \alpha_2 \beta_1).
\end{eqnarray}
Correspondingly, the Bogoliubov transformation of the canonical operators is given by 
\begin{align}
\label{bog}
{b}_p^\dag= \gamma\, a_p^\dagger - \delta a_{-p}
\end{align}
and we may expand the field in terms of this second family: 
\begin{align}
\label{ab}
\varphi_{\beta}(\vec{x},\eta)=
\int{d^{D-1}p}
\left( u_{\beta,\p}(\x, \eta) \,  {b}_p + u^*_{\beta,\p}(\x, \eta) \,  {b}^\dagger_p\right). 
\end{align}
Obviously, at the algebraic level
\begin{align}
\varphi_{\alpha}(\vec{x},\eta)=\varphi_{\beta}(\vec{x},\eta). \label{abba}
\end{align}
In the following we will call 
$\ket{\alpha}$ and 
$\ket{\beta}$  the vacua\footnote{None of the $\ket{\alpha}$ --- states is a  ground state of the Hamiltonian of the theory under consideration. The Hamiltonian depends on time. The BD state is the ground state of the Hamiltonian only at past infinity.} annihilated by the ${a}_p$ and, respectively, the ${b}_p$ operators. Of course the above Bogoliubov transformation is not implementable, the vacua  $\ket{\alpha}$ and 
$\ket{\beta}$ are not equivalent and the scalar product $\scobka{\beta}{\alpha}$ 
appearing below 
is just a formal expression. 

\subsection{Time-ordered \texorpdfstring{$\alpha-\beta$}{} propagator}

Let us now compute the following time-ordered  correlation  function:
\begin{align}\label{ppp}
G_{\alpha-\beta}(x,y)&=\frac{\bra{\beta}T \varphi_{\beta}\left(\eta_1, \vec{x}\right) \varphi_{\alpha}\left(\eta_2, \vec{y}\right)\ket{\alpha}}{\scobka{\beta}{\alpha}}. 
\end{align}
\textcolor{black}{This expression is formal because $|\alpha\rangle$ and   $|\beta\rangle$ do not belong to the same Fock space. However, we will obtain a finite ratio in this equation for generic $|\alpha\rangle$ and   $|\beta\rangle$.}
By taking into account Eq. (\ref{abba}) we get 
%
\begin{eqnarray}
&& G_{\alpha-\beta}(x,y)=
 \int\frac{ {(\eta_1\eta_2)}^{\frac{D-1}{2}}d^{D-1}pd^{D-1}k}{(2\pi)^{D-1}}
 \frac{\bra{\beta}{b}_p{a}^\dag_k\ket{\alpha}}{\scobka{\beta}{\alpha}}\, \times
 \nonumber \\ && \times \left[ \theta(\eta_2-\eta_1)
e^{i\vec{p}\vec{x}-i\vec{k}\vec{y}} \left[ \beta_1 H_{i\mu}^{(1)}(p\eta_1)+\beta_2 H_{i\mu}^{(2)}(p\eta_1)\right]\ 
\left[ \alpha_1 H_{i\mu}^{(1)}(k\eta_2)+\alpha_2 H_{i\mu}^{(2)}(k\eta_2)\right]^* \right.
 \cr && 
+ \left. \theta(\eta_1-\eta_2)e^{i\vec{p}\vec{y}-i\vec{k}\vec{x}} \left[ \beta_1 H_{i\mu}^{(1)}(p\eta_2)+\beta_2 H_{i\mu}^{(2)}(p\eta_2)\right]\ 
\left[ \alpha_1 H_{i\mu}^{(1)}(k\eta_1)+\alpha_2 H_{i\mu}^{(2)}(k\eta_1)\right]^*\right] .\cr && 
\label{bb0}
\end{eqnarray}
Using $(\ref{bog})$ one finds that:
\begin{align}
\label{gd3}
\frac{\bra{\beta}{b}_p{a}^\dag_k\ket{\alpha}}{\scobka{\beta}{\alpha}}= \frac{1}{\gamma}\, \delta^{D-1}(\vec{p}-\vec{k}).
\end{align}
To calculate the integral in $(\ref{bb0})$ we pass to spherical coordinates\footnote{ 
Recall that for an arbitrary function $f(p)$:
\begin{align}\int\frac{d^{D-1}p}{(2\pi)^{D-1}}e^{i\vec{p}\vec{ x}}f(p)=\frac{1}{(2\pi)^{\frac{D-1}{2}}| \x |^{\frac{D-3}{2}}}\int_0^\infty dp \ p^{\frac{D-1}{2}}J_{\frac{D-3}{2}}(p| \x |)f(p).\notag\end{align}}
and take into account the following  \textcolor{black}{formula \cite{bateman2}}:
\begin{align}\label{intkkj}
\int_0^\infty dp \,p^{\nu+1}  K_\mu(ap)K_\mu(bp) J_\nu(cp)=\frac{\sqrt\pi c^\nu \Gamma(\nu+\mu+1)\Gamma(\nu-\mu+1)}{2^{\frac{3}{2}}(ab)^{\nu+1}}(u^2-1)^{-\frac{1}{2}(\nu+\frac{1}{2})}P^{-\nu-\frac{1}{2}}_{-\frac{1}{2}+\mu}(u).
\end{align}
Here $K_\mu$ is the MacDonald function which is related to the  Hankel functions as follows:
\begin{align*}
H^{(1)}_{i\mu}(z)=\frac{2}{\pi} (e^{-i\pi/2})^{i\mu+1}K_{i\mu}(e^{-\frac{i\pi}2} z),\ \ \ \ \
H^{(2)}_{i\mu}(z)=\frac{2}{\pi} (e^{i\pi/2})^{i\mu+1}K_{i\mu}(e^{i\pi/2}z); 
\end{align*}
$P^{-\nu-\frac{1}{2}}_{\mu-\frac{1}{2}}(z)$ is the associated Legendre function of the first kind,  defined on the complex $z$--plane   cut along the real axis from minus infinity to $z=1$; the parameters are such that $u=\frac{a^2+b^2+c^2}{2ab}$ and  $\text{Re}(a+b)>|\text{Im} \, c|$. Let us apply (\ref{intkkj}) to evaluate for instance the  terms on the RHS of (\ref{bb0}) which are proportional to $\alpha_1^*\beta_1$:
\begin{eqnarray}
&& I_1 = \frac{\alpha_1^*\beta_1}{\gamma} \int\frac{ {(\eta_1\eta_2)}^{\frac{D-1}{2}}d^{D-1}p}{(2\pi)^{D-1}} \   \left[ \theta(\eta_2-\eta_1)e^{i\vec{p}(\vec{x}-\vec{y})} H_{i\mu}^{(1)}(p\eta_1)\,  H_{i\mu}^{(1)}(p\eta_2)^* \right. +  \nonumber \\ && + \theta(\eta_1-\eta_2)e^{i\vec{p}(\vec{y}-\vec{x})}\left.  H_{i\mu}^{(1)}(p\eta_2)\   H_{i\mu}^{(1)}(p\eta_1)^* \right] \cr 
&& =  \frac{4 e^{\pi\mu }\alpha_1^*\beta_1{(\eta_1\eta_2)}^{\frac{D-1}{2}}}{(2\pi)^{\frac{D-1}{2}}\pi^2 \gamma| \x-\y |^{\frac{D-3}{2}}}
 \int{ dp} p^{\frac{D-1}{2}}J_{\frac{D-3}{2}}(p| \x-\y |)
 \  \nonumber \cr &&\times \left[ \theta(\eta_2-\eta_1)
 K_{i\mu}(e^{-\frac{i \pi}{2}-i\epsilon}\eta_1 p)\,  K_{i\mu}( e^{\frac{i \pi}{2}-i\epsilon}\eta_2 p) \right.
 +   \theta(\eta_1-\eta_2)\left.   K_{i\mu}(e^{-\frac{i \pi}{2}-i\epsilon}\eta_2 p)\,  K_{i\mu}( e^{\frac{i \pi}{2}-i\epsilon}\eta_1 p)\right]
\cr  
 \cr && =
 \frac{2  \Gamma(\frac{D-1}{2}+i \mu)\Gamma(\frac{D-1}{2}-i \mu)}{\gamma \, \pi \,  (2\pi)^{\frac{D}{2}}}  \, e^{\pi\mu} \alpha_1^*\beta_1\, (Z_-^2-1)^{-\frac{D-2}{4}}P^{-\frac{D-2}{2}}_{-\frac{1}{2}+i\mu}(-Z_-), \cr 
\end{eqnarray}
where 
\begin{align}
 Z_{\pm}=1+\frac{(\eta_1-\eta_2)^2-\left(\vec{x}-\vec{y}\right)^2}{2\eta_1 \eta_2}\pm i\epsilon
 \end{align}
is the hyperbolic distance. The $i\epsilon$ shifts are such that  
the relation $\text{Re} \ (a+b)>|\text{Im} \  c|$ is satisfied. By computing similarly the other terms we get the following expression for the time ordered propagator:
\begin{multline}
G_{\alpha-\beta}(x,y)= 
\frac{  \Gamma(h_+)\Gamma(h_-)}{2\left(\alpha^*_1\beta_1 e^{\mu\pi}-\alpha^*_2\beta_2 e^{-\mu\pi}\right) \,  (2\pi)^{\frac{D}{2}}} \times  
\cr 
\times \Big[ e^{\pi\mu} \alpha_1^*\beta_1\, (Z_-^2-1)^{-\frac{D-2}{4}}P^{-\frac{D-2}{2}}_{-\frac{1}{2}+i\mu}(-Z_-)   +  e^{-\mu\pi}\alpha^*_2 \beta_2 (Z_+^2-1)^{-\frac{D-2}{4}}P^{-(\frac{D-2}{2})}_{-\frac{1}{2}+i\mu}(-Z_+) 
\\-\alpha_2^*\beta_1 e^{-i \pi\frac{D-1}2}  (Z_-^2-1)^{-\frac{D-2}{4}}P^{-\frac{D-2}{2}}_{-\frac{1}{2}+i\mu}(Z_-) -\alpha_1^*\beta_2 e^{+i \pi\frac{D-1}2}  (Z_+^2-1)^{-\frac{D-2}{4}}P^{-(\frac{D-2}{2})}_{-\frac{1}{2}+i\mu}(Z_+) \Big], \label{hh}
\end{multline}
where:

\begin{align}\label{h}
h_\pm={\frac{D-1}{2}\pm i\mu}.
\end{align}
For generic complex $\alpha$'s and $\beta$'s it depends simultaneously on $Z_\pm$ and is only piecewise analytic. It is invariant only w.r.t. the connected part of the isometry group.

Recall also that while in Minkowski space QFT there is a unique Poincar\'e invariant state of positive energy, in de Sitter space there is a family of invariant states  parametrised by solutions of (\ref{comp}), because the notion of positivity of the energy becomes meaningless. The BD state however is peculiar: it is the only one amenable to a thermal interpretation. We will come back on this point as we explain in the section concerning the static patch.

\subsection{Special cases}

\subsubsection{In-In and In-Out Feynman (T-ordered) propagators}

The In-Out propagator corresponds to the choice  $\alpha_1=\alpha_2$ and $\beta_2=0$:
\begin{align}
\label{inout}
G_{ \text{In-Out}}(Z)=\frac{e^{-i\pi(D-2)}}{(2\pi)^{\frac{D}{2}}}(Z_-^2-1)^{-\frac{D-2}{4}} Q_{-\frac{1}{2}+i\mu}^{\frac{D-2}{2}}(Z_-),
\end{align}
where $Q$ is the associated Legendre function of the second kind.
The In-In propagator corresponds to the choice $\alpha_2=\beta_2=0$:
\begin{align}
\label{inin}
G_{ \text{In-In}}(Z)=\frac{\Gamma(h_+)\Gamma(h_-)}{ 2 (2\pi)^{\frac{D}{2}}}  (Z_-^2-1)^{-\frac{D-2}{4}} P^{-\frac{D-2}{2}}_{-\frac{1}{2}+ i\mu}(-Z_-).
\end{align}
The propagator (\ref{inin}) has maximal analyticity properties. It is related to the propagator on the Euclidean sphere via analytical continuation in a suitable time coordinate or in $Z$. On the other hand, (\ref{inout}) it is also related to the Feynman propagators on the Euclidean anti de Sitter (resp. Minkowskian anti de Sitter) via analytical continuation in the radius of curvature (resp. simultaneously in time and radius of curvature) \cite{Akhmedov:2009ta}. We will use these propagators to study the imaginary contributions to the corresponding effective actions.

\subsubsection{Behaviour of the \texorpdfstring{$\alpha-\beta$}{} propagator at coincident points}

The short distance behaviour of the $\alpha-\beta$ propagator is as follows:

\begin{align}
\label{gd1}
G_{\alpha-\beta}(Z) \approx -\frac{\Gamma\left(\frac{D-2}{2}\right)
}{4\pi^{\frac{D}{2}}\delta^{D-2}}\frac{ \left( \alpha^*_1 \beta_1 e^{\pi\mu}+\alpha_2^*\beta_2e^{-\pi\mu}  \right)}{\left(\alpha^*_1\beta_1 e^{\mu\pi}-\alpha^*_2\beta_2 e^{-\mu\pi}\right)} , \ \  \text{as}  \ \delta \to 0,
\end{align}
where $\delta$ is the Minkowski geodesic distance and $Z\approx 1-\frac{\delta^2}{2}$ as $\delta\to 0$. Although  $\delta\to 0$ corresponds to light-like separation, for shortness we will call this limit the coincidence limit. The limit (\ref{gd1}) is equal to the flat spacetime propagator at coincident points multiplied by a certain  constant. The latter is equal to one only for the In-In and the In-Out  propagators. 

The propagator (\ref{inin}) has only the standard singularity at $Z=1$ while all other $\alpha-\beta$ propagators, including In-Out (\ref{inout}), have an extra singularity at $Z=-1$. 

To evaluate explicitly the coincidence limit for the In-In  propagators we may insert the following integral representation of the Legendre function of the first kind \cite{abraste}
\bear
P^{-\frac{D-2}{2}}_{-\frac{1}{2}+ i\mu} (z) = \frac
{2^{{\frac{1}{2}- i\mu}}(z^2-1)^{{\frac{D-2}{4}}}}
{\Gamma({\frac{D-1}{2}}{- i\mu}) \Gamma({\frac{1}{2}+ i\mu} )}
 \int_0^{\infty}dt\, (z+\cosh t)^{{-\frac{D-1}{2}}{- i\mu}} (\sinh t)^{2i\mu}
\label{8.8.1}
\ear
into Eq. \eqref{inin} 
and then set $Z_- = 1$. 
\textcolor{black}{The integral at the rhs is then divergent for $D\geq 2$ but the divergence can be cured as usual by analytical continuation in $D$.} We get 
\bear
G_{\rm In-In}(Z_-=1) =(4\pi)^{-\frac{D}{2}} \Gamma \left(1 - \frac{D}{2}\right) \, 
\frac{\Gamma \left(\frac{D-1}{2} + i \mu \right) \Gamma \left(\frac{D-1}{2} - i \mu \right)}
{\Gamma \left(\frac{1}{2} + i\mu \right) \Gamma \left(\frac{1}{2} - i\mu\right)}.
\label{coincGininfin}
\ear
This result agrees with \cite{candelasraine,dasdun}.



Similar calculations for the coincidence limit in the In-Out case give
\bear
G_{\rm In-Out}(1) = 
\frac{{\rm e}^{-i\pi \frac{D-2}{2}}}{(4\pi)^{\frac{D}{2}} }
\Gamma\Bigl(1-\frac{D}{2}\Bigr)
\frac{\Gamma \left(i\mu + \frac{D-1}{2}\right)}
{\Gamma \left(i\mu - \frac{D-3}{2}\right)}.
\label{coincGinoutfin}
\ear
Although both coincidence limits  \eqref{coincGininfin} are divergent in $D=4$, their ratio is one :
\bear
\frac{G_{\rm In-Out}(1)}{G_{\rm In-In}(1)} \biggl\vert_{D=4} = 1.
\ear
Now we are ready to discuss the imaginary parts of the propagators.


The divergent imaginary part of the generic $\alpha$--
$\beta$ propagator  does not vanish in even dimensions 
$\text{Im} \  G_{\alpha-\beta}(1)\not = 0$ while in odd dimension $\text{Im} \  G_{\alpha-\beta}(1) = 0$ (more details will be given in Section \ref{glo}). In particular for even $D$
\begin{align}
\label{patchim}
\boxed{
\text{Im} \  G_{ \text{In-Out}}(1) =\frac{(-1)^{\frac{D}{2}}e^{-\pi\mu}\left|\Gamma(\frac{D-1}2+i\mu)\right|^2}{(4\pi)^{\frac{D}{2}}\Gamma(\frac{D}{2})},}
\end{align}
and in odd dimensions:
\begin{align}
\boxed{\text{Im} \  G_{ \text{In-Out}}(1)=0.}
\end{align} For the In-In propagator for any $D$
\begin{align}
\boxed{\text{Im} \  G_{ \text{In-In}}(1)=0.}
\end{align}
In Section \ref{glo}  we give a formal explanation why in odd dimensional de Sitter space the Feynman propagators do not contain imaginary contributions. In the concluding Section we also provide a heuristic argument why the effective action in odd dimensional de Sitter space should be real.

{One could ask at this point, which propagator one should use to investigate particle creation in de Sitter space? We come back to the discussion of this point in the concluding section. For now we remark that}

$$
G_{ \text{In-Out}} \approx \frac{e^{-i\pi\frac{D-2}{2}}\Gamma(h_+)}{(2\pi)^{\frac{D-1}{2}}2^{i\mu+1}\Gamma(i\mu+1)}Z^{-h_+} \sim e^{i m L}, \quad {\rm as} \quad Z\to \infty,
$$ 
and 

$$
G_{ \text{In-In}}\approx -\frac{1}{4(2\pi)^{\frac{D+1}{2}}i\mu}\Big[2^{-i\mu}\Gamma(h_+)\Gamma(1-i\mu)Z^{-h_+}-2^{i\mu}\Gamma(h_-)\Gamma(1+i\mu)Z^{-h_-}\Big] \sim A_+ e^{imL} + A_- e^{-imL}, 
$$
as $Z\to \infty$. Here $L$ is the geodesic distance and $Z\sim e^L$, when $L\to \infty$.
{ The Feynman propagator $G_{ \text{In-Out}}$ has the expected large distance behaviour while the $G_{ \text{In-In}}$ does not  \cite{Polyakov:2007mm}, \cite{Akhmedov:2009ta}.}

\section{Pair creation in D=4 de Sitter Space}
\label{paircreation}

As an application of our results on Feynman propagators
in the Poincar\'e patch, let us discuss pair creation
in four-dimensional de Sitter space. Here we shall assume that the usual relation holds between
the pair creation rate $P$ (in the limit of low pair creation) and the imaginary part of the effective Lagrangian,

\bear
P \approx 2 \Im{\cal L}.
\label{PtoL}
\ear
To get the effective Lagrangian, we will use 
\eqref{GtoL} with the In-Out Green's function
at coincident points as given in 
\eqref{coincGinoutfin}, renormalized in $D=4$. 

We perform the renormalization using dimensional regularization and minimial subtraction
in the $\overline{MS}$ scheme. Using the relation 
$
\Gamma(x+\epsilon) = \Gamma(x)(1+\epsilon \psi(x)) + O(\epsilon^2)$, where $\psi=(d/dx){\rm ln}\Gamma(x)$ is the digamma function, 
leads to
\bear
G_{\rm In-Out}^{\rm ren}(1) = \frac{m^2-2}{(4\pi)^2} \biggl\lbrack
-\frac{m^2}{m^2-2} +2\psi\biggl(i\sqrt{m^2-\frac{9}{4}}+\frac{1}{2}\biggr)-i\pi 
\biggr\rbrack.
\label{Ginoutrenfin}
\ear
As expected, this expression has an imaginary part, and using the identity \cite{abraste}
\bear
{\rm Im} \biggl\lbrack \psi\Bigl(\frac{1}{2} + iy\Bigr)\biggr\rbrack = \frac{\pi}{2}\,{\rm tanh} (\pi y),
\ear
we can write it as
\bear
{\rm Im} G_{\rm In-Out}^{\rm ren}(1) = \frac{m^2-2}{(4\pi)^2} \pi\biggl\lbrack {\rm tanh} \biggl (\pi \sqrt{m^2-\frac{9}{4}}  \biggr)-1\biggr\rbrack.
\label{ImGinout}
\ear
Using \eqref{GtoL}
we get the imaginary part of the effective Lagrangian 
in the form

\bear
{\rm Im} {\cal L} = \frac{1}{16\pi}\int_{\infty}^{ m^2}d\bar m^2 (\bar m^2-2)\biggl\lbrack {\rm tanh} \biggl (\pi \sqrt{\bar m^2-\frac{9}{4}}  \biggr)-1\biggr\rbrack.
\ear
Although this integral can be evaluated analytically in terms of polylogarithms, let us discuss here only the large-mass/weak-curvature limit. In this approximation, \eqref{ImGinout} simplifies to 

\bear
{\rm Im} G_{\rm In-Out}^{\rm ren}(1) \approx - \frac{m^2}{8\pi} \,{\rm e}^{-2\pi m},
\label{ImGinoutlargem}
\ear
and integration yields

\bear
{\rm Im} {\cal L} \approx \frac{1}{8\pi^2} m^{\frac{3}{2}}  \,{\rm e}^{-2\pi m}.
\ear
This is in agreement with the pair production rate predicted by the Bogoliubov transformation method 
\cite{lapedes,Mottola:1984ar,Allen:1985ux,Bousso:2001mw}.

This makes it also clear that a previous failure
by Das and Dunne \cite{dasdun} to find the usual relation between the pair creation rate and the imaginary part of the
effective Lagrangian was due to their inappropriate use of the In-In Green's function for the construction of the effective Lagrangian. 
The use of the In-Out Green's function preserves the analogy between the de Sitter and the prototypical constant electric field case. This extends also to another 
aspect of Schwinger pair creation analyzed in \cite{dasdun} for the de Sitter case, namely the possibility of constructing the imaginary part of the effective Lagrangian by a Borel summation of the weak-field expansion of its real part \cite{zinnjustin}. In QED, this can be seen as a natural extrapolation of the optical theorem to 
zero-energy photons, and has been found particularly useful for multiloop considerations 
\cite{dunneschubert,huetrauschschubert}. To see that it works for the case at hand, note 
once more that the weak-field expansion
is equivalent to the large-mass expansion, and that the leading terms of this asymptotic expansion in \eqref{Ginoutrenfin} come from the digamma function. The asymptotic
expansion of its real part is \cite{abraste}

\bear
{\rm Re}  \biggl\lbrack \psi\Bigl(\frac{1}{2} + i\mu \Bigr)\biggr\rbrack \sim {\rm ln}\mu + \sum_{n=1}^{\infty} (-1)^n \frac{1-2^{1-2n}}{2n} B_{2n} \mu^{-2n},
\ear
where the $B_{2n}$ are Bernoulli numbers. Approximating these numbers by their leading asymptotic growth

\bear
B_{2n} \sim (-1)^{n+1} 2 \frac{(2n)}{(2\pi)^{2n}} 
\ear
the series turns into a non-alternating divergent one.
Its Borel summation leads to an imaginary part \cite{dasdun}

\bear
{\rm Im}  \biggl\lbrack \sum_{n=1}^{\infty} (-1)^n \frac{1-2^{1-2n}}{2n} B_{2n} \mu^{-2n} \biggr\rbrack \sim -\pi {\rm e}^{-2\pi \mu} \sim -\pi {\rm e}^{-2\pi m} 
\ear
in agreement with \eqref{ImGinoutlargem}. By integration in $m^2$ one obtains the same correspondence for $\cal L$ itself. 

\newpage

\section{Time-ordered 
propagators in global dS space}
\label{glo}
\subsection{Free modes in global de Sitter space}
\label{formula11}

In global spherical coordinates the de Sitter metric takes the following form:

\begin{align}
ds^2=-dt^2+\cosh^2(t)d\Omega^2.
\end{align}
$d\Omega^2$ is the line element on the unit sphere, and $\sqrt{g}=\cosh^{D-1}(t)\sqrt{|g_{\Omega}|}$, where $|g_{\Omega}|$ is the determinant of the spherical metric. 
Correspondingly,  the KG equation is as follows:
\begin{align*}
(\Box +m^2 )\phi \  = \  \partial_t^2 \varphi +(D-1)\tanh(t) \partial_t-\cosh^{-2}(t) \bigtriangleup_{\Omega}\varphi +m^2\varphi = 0 .
\end{align*}
Here $\bigtriangleup_{\Omega}$ is the Laplace operator on the unit sphere. 
To find the general solution one can  expand $\varphi=\sum_{j,m}\varphi_j(t) Y_{jm}(\Omega)$  in hyperspherical harmonics:
\begin{align*}
\bigtriangleup_{\Omega}Y_{jm}=-j(j+D-2)Y_{jm}
\end{align*}
and get 
\begin{align}
\label{kgeqn}
\bigg(\partial_t^2+(D-1)\tanh(t) \partial_t+j(j+D-2)\cosh^{-2}(t) +m^2\bigg)\varphi_j(t)=0,
\end{align}
where $j$ is a non-negative integer and $m = (1,2,..,N_{j,D})$ ($N_{j,D}$ is the dimension of the $j$--th space of $(D-1)$--dimensional hyperspherical harmonics). 
Two linearly independent solutions are the Ferrers functions $\mathsf{P}$ and $\mathsf{Q}$ also known as the Legendre functions on the cut:
\begin{align}
\label{sol}
\varphi_j( t)=C_1 \cosh^{-\frac{D-1}{2}}(t) \mathsf{P}_{j+\frac{D-3}{2}}^{-i \mu}(\tanh t)+C_2 \frac{2}{\pi}\cosh^{-\frac{D-1}{2}}(t) \mathsf{Q}_{j+\frac{D-3}{2}}^{-i \mu}(\tanh t).
\end{align}
$\mathsf{P}(z)$ and $\mathsf{Q}(z)$ are proportional to the Legendre functions  $P(z)$ and  $Q(z)$ both in the upper and, separately, in the lower complex plane with coefficients such that the Ferrers function are analytic in the cut--complex plane   $\{{\mathbb C}\setminus   ((-\infty -1]\cup[1,\infty))\}$ while the Legendre functions are analytic  in the cut--complex plane   $\{{\mathbb C}\setminus  (-\infty,1]\}$. 

Our goal is again to find In- and Out- modes. 
At  future  infinity
\begin{align}
\mathsf{P}_\nu^{-i \mu}(\tanh t)\approx\frac{e^{-i\mu t}}{\Gamma(1-i \mu)},\   \ \ t\to + \infty,
\end{align}
behaves as a single wave with frequency equal to $\mu$. Modes (\ref{sol}) with $C_2=0$ are usually referred to as Out-modes in global dS space.

As regards the In-modes\footnote{One may use the following known relations:
\begin{align*}
\frac{\sin\left((\nu-i \mu)\pi\right)}{\Gamma\left(\nu+i\mu+1\right)}\mathsf{P}^{%
i \mu}_{\nu}\left(x\right)=\frac{\sin\left(\nu\pi\right)}{\Gamma\left(\nu-i \mu+1%
\right)}\mathsf{P}^{-i \mu}_{\nu}\left(x\right)-\frac{\sin\left(i \mu\pi\right)}{%
\Gamma\left(\nu-i \mu+1\right)}\mathsf{P}^{-i \mu}_{\nu}\left(-x\right),
\end{align*}
\begin{align*}
\frac{2\sin\left(i\mu\pi\right)}{\pi\Gamma\left(\nu-i\mu+1\right)}\mathsf{Q}^{-%
i\mu}_{\nu}\left(x\right)=\frac{1}{\Gamma\left(\nu+i\mu+1\right)}\mathsf{P}^{i\mu%
}_{\nu}\left(x\right)-\frac{\cos\left(i\mu\pi\right)}{\Gamma\left(\nu-i\mu+1%
\right)}\mathsf{P}^{-i\mu}_{\nu}\left(x\right).
\end{align*}},
at  past infinity $\mathsf{P}$ and $\mathsf{Q}$ behave as follows:
\begin{align}
\label{formula4}
\mathsf{P}_\nu^{-i \mu}(\tanh t)\approx\frac{\sin\left(\nu\pi\right)}{\sin\left(i\mu\pi\right) \Gamma(1-i\mu)}e^{i\mu t}-\frac{\sin\left((\nu-i \mu)\pi\right) \Gamma\left(\nu-i\mu+1\right)}{\Gamma\left(\nu+i\mu+1\right)\sin\left(i\mu\pi\right)\Gamma(1+i\mu)}e^{-i\mu t},
\end{align}
\begin{align}
\label{formula5}
\mathsf{Q}_\nu^{-i\mu}(\tanh t)=\frac{\pi \cos (\nu\pi)}{2\sin (i\mu\pi)\Gamma(1-i\mu)}e^{i\mu t}-\frac{\pi \cos((\nu-i\mu)\pi)\Gamma(\nu-i\mu+1)}{2 \sin(i\mu\pi)\Gamma(\nu+i\mu+1)\Gamma(1+i\mu)}e^{-i\mu t}.
\end{align}
To set the coefficient of $e^{ i \mu t}$ in $(\ref{sol})$ to zero one should impose the condition
\begin{align}
\label{formula6}
 C_1 \sin\bigg[\bigg(j+\frac{D-3}{2}\bigg) \pi\bigg]+C_2  \cos\bigg[\bigg(j+\frac{D-3}{2}\bigg)\pi\bigg]=0;
\end{align}
the corresponding solution $(\ref{sol})$ behaves as a single wave at past infinity, usually referred to as In-modes in global dS space.
One sees that in even dimensions $C_1 = 0$ and in odd dimensions $C_2 = 0$. So, in odd dimensions In and Out modes are identical\footnote{This means that the eq. \eqref{kgeqn} for odd $D$ has an integrable (non-scattering) ``potential'', if considered as a quantum mechanical equation. Namely a single wave on one side of the potential passes through it without scattering.}
and this implies that there is no imaginary contribution to the effective action in odd dimensional de Sitter spacetime. Because of that from now on we will not consider $\alpha$--states in odd dimensions. The discussion of $\alpha$--states below  considers only even dimensional spacetimes. None of the $\alpha$--states however diagonalizes the Hamiltonian. Here the situation is different from the one seen in the Poincar\'e patch. There every mode experiences an infinite blue shift towards past infinity such that the modes are almost  ``insensitive'' to the curvature of the de Sitter space and  behave as if they were  in flat space. This means that at past infinity of the Poincar\'e patch the background field is effectively switched off and the Hamiltonian can be diagonalized there.

\subsection{Commutation relations}

Consider the field operator ($\tilde{t} \equiv \tanh t $):
\begin{multline}
\label{phia}
{\varphi}(t,\vec{x})=\sum_{j,m}\cosh(t)^{-\frac{D-1}{2}}\bigg[\bigg(\gamma_1\mathsf{P}_\nu^{-i\mu}(\tilde{t})+\gamma_2 \frac{2}{\pi} \mathsf{Q}_\nu^{-i\mu}(\tilde{t})\bigg) Y_{jm}(\vec{x}){a}_{j,m} +\\+
\bigg(\gamma_1^*\mathsf{P}_\nu^{i\mu}(\tilde{t})+\gamma_2^* \frac{2}{\pi}\mathsf{Q}_\nu^{i\mu}(\tilde{t})\bigg) Y_{jm}^*(\vec{x}){a}_{j,m}^\dagger \bigg],
\end{multline}
where $\vec{x}$ is a unit vector  on the $(D-1)$-dimensional sphere and $[{a}_{j,m},{a}_{j',m'}^\dagger]=\delta_{j,j}\delta_{m,m'}$. 
Let us define $f_{j}=\bigg(\gamma_1\mathsf{P}_\nu^{-i\mu}(\tilde{t})+\gamma_2 \frac{2}{\pi} \mathsf{Q}_\nu^{-i\mu}(\tilde{t})\bigg)\cosh(t)^{-\frac{D-1}{2}}$.   
The canonical commutation relations are 
\begin{align}
\label{formula8}
[\varphi(t,\vec{x}),\dot{\varphi}(t,\vec{y})]=\frac{i \delta(\vec{x}-\vec{y})}{\sqrt{g}} =\sum_{j,m}\bigg(f_{j}\dot{f}^*_{j} Y_{j,m}(\vec{x})Y_{j,m}^*(\vec{y})-f_{j}^*\dot{f}_{j} Y_{j,m}^*(\vec{x})Y_{j,m}(\vec{y})\bigg).
\end{align}
One can change the summation over $m$ in such a way that $ Y_{j,m}^*(\vec{x})Y_{j,m}(\vec{y}) \to  Y_{j,m}(\vec{x})Y_{j,m}^*(\vec{y})$:
\begin{align*}
[\varphi(t,\vec{x}),\dot{\varphi}(t,\vec{y})]=\sum_{j,m}Y_{j,m}(\vec{x})Y_{j,m}^*(\vec{y})\bigg(f_{j}\dot{f}^*_{j}-f_{j}^*\dot{f}_{j} \bigg)=\sum_{j,m}Y_{j,m}(\vec{x})Y_{j,m}^*(\vec{y})W_t(f_{j},f_{j}^*),
\end{align*}
where $W_t$ is the Wronskian of two solutions of $(\ref{kgeqn})$, which does not depend on  $j$: 
\begin{align*}
W_t(f,f^*)=C e^{-\int (D-1)\tanh(t)dt}=C \cosh^{-(D-1)}(t)=\frac{C\sqrt{g_\Omega}}{\sqrt{|g|}},
\end{align*}
where $C$ is some constant which depends on $\gamma_{1,2}$. 
By using the completeness of the hyperspherical harmonics one sees that it shoud be $C=i$. 
Therefore 
\begin{multline}
\label{commutator}
 W_t\bigg(\gamma_1\mathsf{P}_\nu^{-i\mu}(\tilde{t})+\gamma_2 \frac{2}{\pi} \mathsf{Q}_\nu^{-i\mu}(\tilde{t}),\gamma_1^*\mathsf{P}_\nu^{i\mu}(\tilde{t})+\gamma_2^* \frac{2}{\pi} \mathsf{Q}_\nu^{i\mu}(\tilde{t})\bigg)=\\=
\bigg(|\gamma_1|^2  +|\gamma_2|^2 \bigg) \frac{2i \sinh(\mu \pi)}{\pi}-\bigg(\gamma_1^*\gamma_2-\gamma_2^* \gamma_1 \bigg) \frac{2 \cosh(\mu \pi)}{\pi}=i.
\end{multline}
This  condition on the coefficients  $\gamma_{1,2}$ guarantees the canonical commutation relations.
The Out-modes correspond to:
\begin{align}
\gamma_1 = \sqrt{\frac{\pi}{2 \sinh(\mu \pi)}}, \ \ \text{and} \ \ \gamma_2 = 0.
\end{align}
The In-modes correspond to:
\begin{eqnarray*}
\gamma_1 = \sqrt{\frac{\pi}{2 \sinh(\mu \pi)}}, \ \ \ \ \gamma_2 = 0, \ \ \ \ \text{in odd dimensions},\\
\text{and} \ \ \ \gamma_2 = \sqrt{\frac{\pi}{2 \sinh(\mu \pi)}}, \ \ \ \ \gamma_1 = 0, \ \ \ \ \text{in even dimensions}.
\end{eqnarray*}

\subsection{Bogoliubov transformation}

As before, let us consider a second mode expansion of the field operator $\varphi(t,\vec{x})$ of the same form as (\ref{phia}) but with other coefficients $\chi_1$ and $\chi_2$ and corresponding operators ${b}^\dagger_{j,m}$ and ${b}_{j,m}$.  $\chi_1$ and $\chi_2$ also obey to the  relation 
$(\ref{commutator})$. 
Using $(\ref{formula5})$ 
and comparing the two expressions of the field operator 
we obtain two identities:
\begin{eqnarray*}
\gamma_2 Y  C_+ {a}+(\gamma_1^* +\gamma_2^*  C_-^*) Y^*  {a}^\dagger=\chi_2 Y  C_+ {b} + (\chi_1^* +\chi_2^*  C_-^*) Y^*  {b}^\dagger,\\
 \gamma_2^* Y^*  C_+^* \hat{a}^\dagger+(\gamma_1 +\gamma_2 C_-) Y  {a} =\chi_2^* Y^*  C_+^* {b}^\dagger+(\chi_1 +\chi_2 C_-) Y  {b} ,
\end{eqnarray*}
we denote $Y_{j,m}=Y$ here to simplify these expressions and
\begin{align*}
C_+=-i\frac{ \Gamma(\nu-i\mu+1)}{\sinh (\mu \pi) \Gamma(\nu+i\mu+1)}\ \ \ \text{and} \ \ \ C_-=i \coth(\mu \pi).
\end{align*}
One can write the solution of the above equations in the following form $\hat{a}^\dagger=u_b \hat{b}^\dagger + u_a \hat{a}$.  For the discussion below we need to know only $u_b$:
\begin{align}
\label{ub}
u_b=\frac{|\chi_1+\chi_2 C_-|^2-|C_+|^2|\chi_2|^2}{\gamma_2^*\chi_2+\gamma_1^*\chi_1+\gamma_2^*\chi_1C_-^*+\gamma_1^*\chi_2C_-}.
\end{align}
Note that $u_b$ does not depend on $j$.

\subsection{Feynman  \texorpdfstring{$\gamma-\chi$}{} propagator}

\subsubsection{Mode expansion}
Here we compute the Feynman propagator between $\ket{\gamma}$ and $\ket{\chi}$ states, which are defined\footnote{Note that these $\gamma$--states are not the same ones as defined in the section on the Poincar\'e patch. We discuss the relation between these $\gamma$--states and those in the Poincar\'e patch at the end of this section.} as  ${a}_{j,m}\ket{\gamma}=0$ and ${b}_{j,m}\ket{\chi}=0$:
Let us denote 
\begin{eqnarray}
f_{1,j}(t_1)=\bigg(\gamma_1\mathsf{P}_\nu^{-i\mu}(\tilde{t}_1)+\gamma_2 \frac{2}{\pi} \mathsf{Q}_\nu^{-i\mu}(\tilde{t}_1)\bigg)\cosh(t_1)^{-\frac{D-1}{2}},  \nonumber \\ f_{2,j}(t_2)=\bigg(\chi_1\mathsf{P}_\nu^{-i\mu}(\tilde{t}_2)+\chi_2 \frac{2}{\pi} \mathsf{Q}_\nu^{-i\mu}(\tilde{t}_2)\bigg)\cosh(t_2)^{-\frac{D-1}{2}} .
\end{eqnarray}
For  $t_2<t_1$ we get
\begin{eqnarray}
G_{\gamma-\chi}(t_1,\vec{x}|t_2,\vec{y}) = \frac{ \bra{ \chi}T\varphi(t_1,\vec{x})\varphi(t_2,\vec{y})\ket{\gamma}}{\scobka{\chi}{\gamma}} = 
\sum_{j,m}f_{2,j}(t_1)f_{1,j}^*(t_2)u_b Y_{j,m}(\vec{x}) Y_{j,m}^*(\vec{y}),
\end{eqnarray}
where $u_b$ is defined in Eq. (\ref{ub}).
 When $D >2$ one can use the relation:
\begin{align*}
\sum_{m=1}^{N_{j,D}} Y_{j,m}(\vec{x})Y_{j,m}^*(\vec{y})=\frac{2 j + D - 2}{|S^{D-1}| (D-2)}C_{j}^{\frac{D-2}{2}}(\vec{x}\cdot \vec{y}),
\end{align*}
where $C_{j}^{\frac{D-2}{2}}(\vec{y} \vec{x})$ are the Gegenbauer polynomials and $|S^{D-1}|$ is the volume of the $(D-1)$-dimensional sphere:
\begin{multline}
\label{propalphabeta}
G_{\gamma-\chi}(t_1,\vec{x}|t_2,\vec{y})=\frac{u_b \cosh(t_2)^{-\frac{D-1}{2}}\cosh(t_1)^{-\frac{D-1}{2}}}{|S^{D-1}| (D-2)}\sum_j\bigg(2 j + D - 2\bigg)\times\\ \times \bigg(\chi_1\mathsf{P}_{j+\frac{D-3}{2} }^{-i\mu}(\tilde{t_1})+\chi_2 \frac{2}{\pi} \mathsf{Q}_{j+\frac{D-3}{2} }^{-i\mu}(\tilde{t_1})\bigg)  \bigg(\gamma_1^*\mathsf{P}_{j+\frac{D-3}{2} }^{i\mu}(\tilde{t_2})+\gamma_2^* \frac{2}{\pi} \mathsf{Q}_{j+\frac{D-3}{2} }^{i\mu}(\tilde{t_2})\bigg)C_{j}^{\frac{D-2}{2}}(\vec{y}\vec{x})=
\end{multline}
\begin{multline}
\label{fullab}=\frac{2u_b}{\pi}\Big( A1 \Big) \Big[\chi_1 \gamma_2^* + i \chi_1 \gamma_1^* \coth\mu\pi\Big]-\Big( A1 \Big)^*\frac{2u_b \chi_2 \gamma_1^*}{\pi \sinh^2 \mu\pi} -\\-
\Big( A2 \Big)\frac{2u_b}{\pi \sinh \mu\pi}\Big[ i \chi_1\gamma_1^*-\chi_2 \gamma_1^* \coth\mu\pi\Big] +\Big( A3 \Big)\frac{4u_b}{\pi^2} \Big[ \chi_2 \gamma_2^*+i \chi_2 \gamma_1^* \coth \mu\pi\Big]=
\end{multline}
\begin{multline}
\label{abfullglob}
=\frac{(-1)^\frac{D-2}{2}}{2(2\pi)^\frac{D}{2}} \bigg[ S^\frac{D-2}{2}_{i\mu-\frac{1}{2}}(Z_+)\Big(B_1+B_2e^{\mu\pi}-\frac{i\pi}{2}B_3\Big)+S^\frac{D-2}{2}_{i\mu-\frac{1}{2}}(Z_-)\Big(B_1+B_2e^{-\mu\pi}+\frac{i\pi}{2}B_3\Big)+\\+B_4\Big(S^\frac{D-2}{2}_{-i\mu-\frac{1}{2}}(Z_+)+S^\frac{D-2}{2}_{-i\mu-\frac{1}{2}}(Z_-)\Big)+B_3 \frac{i\pi^2}{2\cosh \mu\pi} \Big(F^\frac{D-2}{2}_{i\mu-\frac{1}{2}}(Z_+)-F^\frac{D-2}{2}_{i\mu-\frac{1}{2}}(Z_-)\Big) \bigg].
\end{multline}
In the intermediate step  $\eqref{A1}, \eqref{A2}$ and $\eqref{A3}$ are the expressions computed in Appendix \ref{appendix} (recall that $D$ is assumed to be  even); 
\begin{align}
\label{ZZ}
Z_\pm \equiv Z\pm i \epsilon =\frac{-\tilde{t_1}\tilde{t_2}+\vec{x}\vec{y}}{\sqrt{1-\tilde{t_1}^2}\sqrt{1-\tilde{t_2}^2}}\pm i\epsilon.
\end{align}
is again the scalar invariant (hyperbolic distance) between  $(t_1,\vec{x})$ and $(t_2,\vec{y})$ expressed in global de Sitter coordinates. We also used the following notations:
\begin{align}
\label{FS}
 F_a^k(x)=(x^2-1)^{-\frac{k}{2}} P_a^k(x),  \quad {\rm and} \quad S_a^k(x)=(x^2-1)^{-\frac{k}{2}} Q_a^k(-x),
\end{align}
\begin{align*}
B_1 & =\frac{2 u_b}{\pi}\Big[\chi_1 \gamma_2^* +i \chi_1 \gamma_1 ^* \coth\mu\pi\Big],  & \ \  B_4 & =-\frac{2 u_b}{\pi} \frac{\chi_2 \gamma_1^*}{\sinh^2 \mu\pi}, \\
B_2 & =-\frac{2 u_b}{\pi \sinh \mu\pi} \Big[ i \chi_1 \gamma_1^*-\chi_2 \gamma_1^* \coth\mu\pi\Big],  &  B_3 & =\frac{4 u_b}{\pi^2}\Big[ \chi_2 \gamma_2^*+i \chi_2 \gamma_1^* \coth \mu\pi\Big]. \end{align*}

\subsection{Behavior of the generic  \texorpdfstring{$\gamma-\chi$}{} propagator at coincident points}

As we did in the Poincar\'e patch, let us expand 
$
Z\approx 1-\frac{\delta^2}{2} \ \ \text{where} \ \ \delta \to 0.
$
The coefficient of $B_3$ in $\eqref{abfullglob}$ vanishes.
To find the behavior of the other terms we need the following formula \cite{Bateman}:
\begin{align}
\label{Sdef}
S^\frac{D-2}{2}_{i\mu-\frac{1}{2}}(Z_\pm)=\frac{|\Gamma(h_+)|^2}{2^{\frac{D}{2}} \Gamma\big(\frac{D}{2}\big)}\bigg[\mp i(-1)^{\frac{D}{2}}e^{\mp \mu\pi} F\Big(\frac{1+Z_\pm}{2}\Big)+F\Big(\frac{1-Z_\pm}{2}\Big)\bigg],
\end{align}
where $h_\pm$ are defined in $\eqref{h}$ and 
$
F(x) \equiv {}_2F_1 \Big(h_+,h_-,\frac{D}{2},x \Big).
$
Only the first term in $\eqref{Sdef}$ is singular at $Z=1$. So in the limit in question one can divide the propagator into two parts:
\begin{align}
G_{\gamma-\chi}\Big( Z\approx 1-\frac{\delta^2}{2} \Big) =G_{\text{sing}}(\delta)+G_{\text{finite}}(\delta).
\end{align}
Since 
\begin{align}
F\Big(\frac{1+Z_\pm}{2}\Big) \approx -\frac{1}{4} \frac{\Gamma\big(\frac{D}{2}\big)\Gamma\big(\frac{D-2}{2}\big)}{|\Gamma(h_+)|^2} \frac{4^\frac{D}{2}}{\delta^{D-2}} , \ \ \ \text{as} \ \ \delta\to 0,
\end{align}
one finds that:
\begin{align}
\label{GP}
G_\text{sing}(\delta)\approx\frac{\Gamma\Big(\frac{D-2}{2}\Big)}{4 \pi^\frac{D}{2} \delta^{D-2}}\frac{2 u_b}{\pi} \bigg[ \cosh \mu\pi \Big(\chi_1 \gamma_1^*+\chi_2 \gamma_2^* \Big)-i \sinh \mu\pi \Big(\chi_1 \gamma_2^*-\chi_2 \gamma_1^* \Big)\bigg] \notag =\\=
\frac{\Gamma\Big(\frac{D-2}{2}\Big)}{4 \pi^\frac{D}{2} \delta^{D-2}} T, \  \ \text{as} \  \delta\to 0,
\end{align}
which is the same behavior of the flat space propagator at coincident points multiplied by some constant $T$ which depends on $\gamma$ and $\chi$.
The finite term in $\eqref{Sdef}$ is as follows:
\begin{align}
\label{GC}
G_\text{finite} \approx\frac{2 (-1)^\frac{D-2}{2} |\Gamma(h_+)|^2 u_b}{\pi(4\pi)^\frac{D}{2} \Gamma\Big(\frac{D}{2} \Big)} \bigg[\chi_1 \gamma_2^*+\chi_2 \gamma_1^* \bigg], \ \ \text{as} \ \delta\to 0.
\end{align}
Below we study both contributions $G_\text{sing}$ and $G_\text{finite}$ for special values of $\gamma_{1,2}$ and $\chi_{1,2}$. As in the case of the Poincar\'e patch the generic $\gamma-\chi$ propagator has another singularity at $Z=-1$. Moreover, there can be a divergent imaginary part.

\subsection{Specially interesting  cases}

Because in odd dimensions $\ket{\text{Out}}=\ket{\text{In}}$ as we explained above, in such a case $G_{\text{In-In}}=G_{\text{In-Out}}$ and the first one is always real at the coincident points.

\subsubsection{In-In in even dimensions}

According to the definition of the In-modes, the In-In propagator corresponds to 
\begin{align*}
\gamma_1=\chi_1 = \sqrt{\frac{\pi}{2 \sinh(\mu \pi)}}, \ \ \ \  \gamma_2=\chi_2 = 0, \ \ \  \ \ \ u_b=1,
\end{align*}
as follows from Eq. (\ref{ub}). Hence
\begin{align}
G_\text{In-In}(\delta)\approx \frac{\Gamma\Big(\frac{D-2}{2}\Big)}{4 \pi^\frac{D}{2} \delta^{D-2}} \coth \mu\pi.
\end{align}
This propagator has no imaginary part and its finite part vanishes.

\subsubsection{In-Out in even dimensions}
Here
\begin{align*}
\gamma_2=\chi_1 = \sqrt{\frac{\pi}{2 \sinh(\mu \pi)}}, \ \ \ \  \gamma_1=\chi_2 = 0 , \quad  \ \ u_b = i\tanh(\mu\pi).
\end{align*}
The sum of $\eqref{GP}$ and $\eqref{GC}$ has the following form:
\begin{align}
G_\text{In-Out}(\delta)\approx \frac{\Gamma\Big(\frac{D-2}{2}\Big)}{4 \pi^\frac{D}{2} \delta^{D-2}} \tanh \mu\pi+i\frac{  (-1)^\frac{D-2}{2} |\Gamma(h_+)|^2 }{(4\pi)^\frac{D}{2} \Gamma\Big(\frac{D}{2} \Big) \cosh \pi\mu}, \quad  \ \ \delta \to 0.
\end{align}
This propagator has a finite imaginary part:
\begin{align}
\label{globalim}
\boxed{
\text{Im} \ G_\text{In-Out}(Z=1)=\frac{(-1)^\frac{D-2}{2} |\Gamma(h_+)|^2 }{(4\pi)^\frac{D}{2} \Gamma\Big(\frac{D}{2} \Big) \cosh\pi\mu}.}
\end{align}
This result corresponds to the one obtained for the Poincar\'e patch in $\eqref{patchim}$.  But there is an important difference, because the In--state in global de Sitter space does not coincide with the In-- (or BD) state in the Poincar\'e patch. As a consequence the expression $\eqref{globalim}$ differs from $\eqref{patchim}$.

\subsubsection{ \texorpdfstring{$\gamma-\chi$}{} propagator}

In this case $
\gamma_1=\chi_1, \ \ \ \ \gamma_2=\chi_2, \ \  \ \ u_b=1;
$
\begin{align}
G_{\gamma-\gamma}\approx\frac{\Gamma\Big(\frac{D-2}{2}\Big)}{4 \pi^\frac{D}{2} \delta^{D-2}}\frac{2}{\pi} \bigg[ \cosh \mu\pi \Big(|\gamma_1|^2 +|\gamma_2|^2 \Big)+i \sinh \mu\pi \Big(\gamma_1^* \gamma_2-\gamma_2^* \gamma_1 \Big)\bigg]= \nonumber \\=
\frac{\Gamma\Big(\frac{D-2}{2}\Big)}{4 \pi^\frac{D}{2} \delta^{D-2}} T, \  \ \text{as} \  \delta\to 0,
\end{align}
The condition $\eqref{commutator}$ implies that:
$
T \ge 1.
$
The minimum value of $T$ correspond to 
\begin{align}
-i\gamma_1=\gamma_2=\sqrt{\frac{\pi}{4e^{\mu\pi}}}.
\end{align}
The value $T=1$ cannot be achieved if $\gamma_1$ and $\gamma_2$ are both real.

\subsection{Relation to the \texorpdfstring{$\alpha$}--modes in the Poincar\'e patch}

Let us make the following transformation:

\begin{alignat*}{10}
\gamma_1 & =\alpha_1^*+\alpha_2^*, & \qquad \gamma_2 & =i\Big( \alpha_2^*-\alpha_1^* \Big), \\
\chi_1 & =\beta_1^*+\beta_2^*, & \chi_2 & =i \Big(\beta_2^*-\beta_1^* \Big),
\end{alignat*}
then the condition $\eqref{commutator}$ for $\gamma_{1,2}$  and $\chi_{1,2}$ transform into:
\begin{align*}
|\alpha_1|^2 e^{\mu\pi}-|\alpha_2|^2 e^{-\mu\pi}=\frac{\pi}{4},
\end{align*}
and the same condition for $\beta_{1,2}$. In this case $u_b$:
\begin{align*}
u_b=\frac{\pi}{4\Big(\alpha_1 \beta_1^* e^{\mu\pi}-\alpha_2 \beta_2^* e^{-\mu\pi}\Big)},
\end{align*}
as follows from $\eqref{ub}$. With the new coefficients $\beta_{1,2}$ and $\alpha_{1,2}$ the finite part of the $\gamma-\chi$ propagator has the following form:
\begin{align*}
G_\text{finite}(\delta) \approx \frac{i (-1)^\frac{D-2}{2} |\Gamma(h_+)|^2 }{(4\pi)^\frac{D}{2} \Gamma\Big(\frac{D}{2} \Big)} \frac{\beta_1^* \alpha_2-\beta_2^*\alpha_1}{\alpha_1 \beta_1^* e^{\mu\pi}-\alpha_2 \beta_2^* e^{-\mu\pi}}, \ \ \text{as} \ \delta\to 0,
\end{align*}
and the singular part is:
\begin{align*}
G_\text{sing}(\delta)\approx\frac{\Gamma\Big(\frac{D-2}{2}\Big)}{4 \pi^\frac{D}{2} \delta^{D-2}}\frac{\beta_1^* \alpha_1 e^{\mu\pi} + \beta_2^* \alpha_2 e^{-\mu\pi}}{\alpha_1 \beta_1^* e^{\mu\pi}-\alpha_2 \beta_2^* e^{-\mu\pi}}=
\frac{\Gamma\Big(\frac{D-2}{2}\Big)}{4 \pi^\frac{D}{2} \delta^{D-2}} T, \  \ \text{as} \  \delta\to 0.
\end{align*}
This is a manifestation of the fact that there is the so called Euclidean state $\ket{\text{E}}$ in global de Sitter \cite{Mottola:1984ar}, \cite{Bousso:2001mw}, which corresponds to the $\ket{\text{BD}}$ or $\ket{\text{In}}$--state in the Poincar\'e patch. Namely BD--BD (or In--In) propagator in 
the Poincar\'e patch coincides with the E--E propagator, which is just the $\gamma-\gamma$--propagator for a concrete value of $\gamma$, in global de Sitter. Of course this coincidence is no surprise since the BD vacuum is maximally analytic \cite{moschella1,moschella2}.

One can prove that for any coefficients $\alpha, \beta$:
\begin{align*}
0<|T|<\infty.
\end{align*}
Let us consider the following values of the coefficients:

\begin{center}
  \begin{tabular}{l  c}
  $
   \alpha_1 =\sqrt{x^2+1} e^{-\frac{\mu\pi}{2}} \sqrt{\frac{\pi}{4}},$ &\ \ \ \ \ \ \ \ \ $ \beta_1 =\sqrt{x^2+1} e^{-\frac{\mu\pi}{2}} \sqrt{\frac{\pi}{4}}, $\\$
    \alpha_2=i x e^{\frac{\mu\pi}{2}} \sqrt{\frac{\pi}{4}},$&$ \beta_2 =x e^{\frac{\mu\pi}{2}} \sqrt{\frac{\pi}{4}}
  
    $
  \end{tabular}
\end{center}
for some $x$. Then the corresponding Feynman $\alpha-\beta$ propagator has a divergent imaginary part, because:

\begin{align}
T=\frac{(x^2+1)+ix^2}{(x^2+1)-ix^2}.
\end{align}
Thus, in the case of generic $\gamma-\chi$--propagator one can have a complex singular part both in the global de Sitter and in the Poincar\'e patch. Moreover, as we have mentioned several times above, the generic $\gamma-\chi$--propagator has another singularity at $Z=-1$.
 
\section{The static patch}
\label{sta}
In this section we shall discuss the properties of a  massive real scalar field in the static patch of the de Sitter manifold. The main feature of this patch is the existence of a time-like Killing vector that allows one to introduce a notion of energy;  the free Hamiltonian is the generator of the above-mentioned time translations. 

For notational simplicity here we restrict ourselves to the two-dimensional spacetime; the following can  easily be generalized to the general case. 
The de Sitter metric now is written as 
\begin{gather}
    ds^2 = \left(1-r^2\right) dt^2 - \frac{dr^2}{1-r^2} = \frac{dt^2 - dx^2}{\cosh^2 x},\quad r=\tanh x.
\end{gather}
In these coordinates the scalar invariant is given by 
\begin{gather}
    Z = \frac{\cosh(t_1 - t_2) + \sinh x_1 \sinh x_2}{\cosh x_1 \cosh x_2}.
\end{gather}
 Time translation invariant Green functions now solve the following equation: 
\begin{gather}
\left[\partial_t^2 - \partial_x^2 + \frac{m^2}{\cosh^2 x}\right] G_F(x,y|t - t')=-i\delta(x-y)\delta(t-t'),\label{sdseq}
\end{gather}
Let us consider the spatial part of this differential equation. The eigen--functions of the operator consideration solve the following equation:

\begin{gather}
\left[- \partial_x^2 + \frac{m^2}{\cosh^2 x}\right] \psi_k(x) = k^2 \psi_k(x)
\end{gather}
Its general solution is a combination of Legendre functions
\begin{gather}
\psi_k(x) = A_k \mathsf{P}^{ik}_{-\frac12+ i\mu}(\tanh x) + B_{k} \mathsf{P}^{-ik}_{-\frac12+ i\mu}(\tanh x)
\end{gather}
This problem is obviously related to the text book quantum mechanical one-dimensional problem with the potential $V(x) = \frac{m^2}{\cosh^2 x}$. The spectrum of such a problem is known. At each energy $E = k^2>0$ there are two states, that can be charactirized by the quantum numbers $k,-k$. It means that there exists a full set of functions satisfying the orthogonality

\begin{gather}
\int dx \mathsf{M}^{i k}_{-\frac12 + i \mu}(x)  \mathsf{M}^{i k'}_{-\frac12 + i \mu}(x) = \frac{2\sinh \pi k}{k} \delta(k+k'),
\end{gather}
and the completness
\begin{gather}
    \int\limits^\infty_{-\infty} \frac{k dk}{2\sinh \pi k} \mathsf{M}^{ik}_{-\frac12+i\mu}(\tanh x)\mathsf{M}^{-ik}_{-\frac12+i\mu}(\tanh y) = \delta(x-y), \label{orthoLegP}
\end{gather}
relations. The exact form of the coefficients $A_k$ and $B_k$ is not important for us here and will be provided elsewhere.

Using these relations one can obtain the following  integral representation of the Green function:
\begin{gather}
    G_F(x,y|t) = \iint^\infty_{-\infty} \frac{d\omega}{2\pi}\frac{k dk}{2\sinh \pi k} A(\omega,k) \mathsf{M}^{ik}_{-\frac12+i\mu}(\tanh x)\mathsf{M}^{-ik}_{-\frac12+i\mu}(\tanh y) e^{-i \omega t}. \label{sdsans}
\end{gather}
The Feynman prescription gives
\begin{gather}
   A(\omega,k) = \frac{-i}{-\omega^2+k^2+i\epsilon},
\end{gather}
precisely as in  flat space-time. After that the integral over $\omega$ in \eqref{sdsans} can be calculated, which yields 
\begin{gather}
   G_F(x,y|t) = \int\limits^\infty_{-\infty} \frac{dk}{4\sinh \pi |k|} \mathsf{M}^{ik}_{-\frac12+i\mu}(\tanh x)\mathsf{M}^{-ik}_{-\frac12+i\mu}(\tanh y) e^{-i |k| |t|}.
\end{gather}
From this one  we can rewrite it via the positive defined harmonics as
\begin{gather} 
   G_F(x,y|t) = \notag\\=
   \int\limits^\infty_{0} \frac{d\omega}{4\sinh \pi \omega} \left(\mathsf{M}^{-i\omega}_{-\frac12+i\mu}(\tanh x)\mathsf{M}^{i\omega}_{-\frac12+i\mu}(\tanh y)+\mathsf{M}^{i\omega}_{-\frac12+i\mu}(\tanh x)\mathsf{M}^{-i\omega}_{-\frac12+i\mu}(\tanh y)\right) e^{-i \omega |t|}.\label{rewrittenFeynman}
\end{gather}
This is the Feynman propagator for the ground state of the free Hamiltonian in the Static patch. 

We will show now that the last expression cannot be a function of the invariant $Z$. In fact, consider the point $x=y=0$. Then,

\begin{gather}
   G_F(0,0|t) = \int\limits^\infty_{-\infty} \frac{dk}{4\sinh \pi |k|} \, \left|\mathsf{M}^{ik}_{-\frac12+i\mu}(0) \right|^2 \, e^{-i |k| |t|}.
\end{gather}
If this were a function of $Z = \cosh t$, then it would have been periodic under the change $t \to t + 2 \pi i$, but if one analytically continues this function in the complex variable $\tau$, he finds that:

\begin{gather}
   G_F(0,0|0) - G_F(0,0|2\pi i) = \int\limits^\infty_{-\infty} \frac{dk}{4\sinh \pi |k|} \, \left|\mathsf{M}^{ik}_{-\frac12+i\mu}(0) \right|^2 \, \left[1 - e^{- 2 \pi |k|}\right] > 0.
\end{gather}
Which means that the ground state of the time independent free Hamiltonian in the Static patch does not respect the de Sitter isometry, i.e. does not provide two--point functions which depend only on the invariant $Z$. So then one can ask which state in Static patch does provide de Sitter invariant propagators?

\textcolor{black}{The well-known answer is that the BD or Euclidean state correspond to the thermal one in the Static patch \cite{gibbons,sewell, moschella1,moschella2} with the inverse temperature equal to $\beta=2\pi$. We give here a new derivation of this fact}. The thermal Wightman propagator in the real time formalism has the following form
\begin{eqnarray}
    G_W(x,y|t) = \int\limits^\infty_{-\infty}  \frac{d\omega}{4\sinh \pi \omega} \, \frac{e^{i\omega t}}{e^{\beta \omega}-1} \,  \Bigl[\mathsf{M}^{i\omega}_{-\frac12+i\mu}(\tanh x) \mathsf{M}^{-i\omega}_{-\frac12+i\mu}(\tanh y) \Bigr. + \nonumber \\ + \Bigl. \mathsf{M}^{i\omega}_{-\frac12+i\mu}(\tanh y) \mathsf{M}^{-i\omega}_{-\frac12+i\mu}(\tanh x)\Bigr], \quad {\rm where} \quad \beta = 2\pi,
\end{eqnarray}
which appears as a modification of \eqref{rewrittenFeynman} where we changed the distribution to the thermal one with $n(\omega) = \frac{1}{e^{\beta\omega}-1}$.
Using the Wightman function one can construct the  other propagators, e.g.:

\begin{eqnarray}
G_R(x,y|t) = \theta(t) \, {\rm Im} G_W(x,y|t), \nonumber \\ G_F(x,y,|t) = \theta(t) \, G_W(x,y|t) + \theta(-t) \, G_W(y,x|-t).
\end{eqnarray}
Then using the identity

\begin{gather}\label{sds:usef1}
    \frac{1}{e^{2\pi \omega}-1} = -\sum_{n= -\infty}^{+\infty} \int\limits^\infty_{-\infty} \frac{d\tau}{2\pi i} \frac{e^{-i\omega \tau}}{\tau+2\pi i n - i \epsilon},
\end{gather}
in the last equation, one obtains that:
\begin{gather}
    G_{W}(x,y|t) = -\sum_n \int\limits^\infty_{-\infty} \frac{d\tau}{2\pi i}\int\limits^\infty_{-\infty} \frac{d\omega}{4\sinh \pi \omega}  \frac{e^{i\omega\left(t-\tau\right)}}{\tau+2\pi i n - i \epsilon} \times \nonumber \\ \times \left[\mathsf{M}^{ik}_{-\frac12+i\mu}(\tanh x)\mathsf{M}^{-ik}_{-\frac12+i\mu}(\tanh y)+\mathsf{M}^{ik}_{-\frac12+i\mu}(\tanh y)\mathsf{M}^{-ik}_{-\frac12+i\mu}(\tanh x)\right]=\notag\\=
   2\sum_n \int\limits^\infty_{-\infty} \frac{d\tau}{2\pi i} \frac{\Im G_{W}(x,y|t-\tau)}{\tau+2\pi i n - i \epsilon} =2 \sum_n \int\limits^\infty_{-\infty} \frac{d\tau}{2\pi i} \frac{\Im G_{W}(x,y|\tau)}{t-\tau+2\pi i n - i \epsilon}, \label{sds:FinalForm}
\end{gather}
where we shifted the integral over the variable $\tau$ in the last line and used the relation
\begin{gather*}
       \Im G_{W}(x,y|t) = \frac{1}{2} \left[G_{W}(x,y|t)-G_{W}^*(x,y|t)\right] =\notag\\=  \int\limits^\infty_{-\infty} \frac{d\omega}{4\sinh \pi \omega} \frac{\sin \omega t}{e^{2\pi \omega}-1} \left[\mathsf{M}^{i\omega}_{-\frac12+i\mu}(\tanh x)\mathsf{M}^{-i\omega}_{-\frac12+i\mu}(\tanh y)+\mathsf{M}^{i\omega}_{-\frac12+i\mu}(\tanh y)\mathsf{M}^{-i\omega}_{-\frac12+i\mu}(\tanh x)\right] = \notag\\
       =-\int\limits^\infty_{-\infty}\frac{d\omega}{8\sinh \pi \omega}  e^{i\omega t}\left[\mathsf{M}^{i\omega}_{-\frac12+i\mu}(\tanh x)\mathsf{M  }^{-i\omega}_{-\frac12+i\mu}(\tanh y)+\mathsf{M}^{i\omega}_{-\frac12+i\mu}(\tanh y)\mathsf{M}^{-i\omega}_{-\frac12+i\mu}(\tanh x)\right].
\end{gather*}
At the end using $\Im G_{W} = \Im G_{BD}$ we obtain that

\begin{gather}
      G_{W}(x,y|t) = 2\sum_n \int\limits^\infty_{-\infty} \frac{d\tau}{2\pi i} \frac{\Im G_{BD}\left(\frac{\cosh(\tau) + \sinh x_1 \sinh x_2}{\cosh x_1 \cosh x_2}\right)}{t-\tau+2\pi i n - i \epsilon}. \label{sds:BDForm}
\end{gather}
Now one can perform the summation over $n$ in the equation \eqref{sds:BDForm} to get
\begin{gather}
    G_{W}(x,y|t) = \int\limits^\infty_{-\infty} \frac{d\tau}{2\pi i} \Im G_{BD}\left(\frac{\cosh(\tau) + \sinh x_1 \sinh x_2}{\cosh x_1 \cosh x_2}\right) \coth\left(\frac{t-\tau - i \epsilon}{2}\right).
\end{gather}
With the change of variables $z=e^\tau$ and
\begin{align*}
    \coth\left(\frac{t-\tau - i \epsilon}{2}\right)=\frac{e^\frac{t-\tau}{2}+e^{-\frac{t-\tau}{2}}}{ e^\frac{t-\tau-i\epsilon}{2}-e^{-\frac{t-\tau-i \epsilon}{2}}}=\frac{z+e^t}{e^t-z-i\epsilon},
\end{align*}
this integral can be rewritten as

\begin{gather}
    G_{W}(x,y|t) =- \int\limits^\infty_0 \frac{dz}{2\pi i z} \Im G_{BD}\left(\frac{z+1/z + 2\sinh x_1 \sinh x_2}{2\cosh x_1 \cosh x_2}\right) \frac{z + e^t}{z-e^t+i\epsilon}.
\end{gather}
We change the contour of integration to be $C=C_+ \cup C_-, C_{\pm} = {\alpha\left(1\pm i \delta\right),\alpha\in \mathbb{R}_+}$

\begin{gather}
    G_{W}(x,y|t) = \frac12\int_C \frac{dz}{2\pi i z} G_{BD}\left(\frac{z+1/z+ 2\sinh x_1 \sinh x_2}{2\cosh x_1 \cosh x_2}\right)\frac{z + e^t}{z-e^t+i\epsilon}.
\end{gather}
This integral can be taken by the Cauchy theorem. In fact, noticing that as $z\to \infty$ the integrand goes to zero as $\sim z^{-\frac32}$, which allows one to close the contour at infinity. The only non-zero contribution comes from the $z=e^t$ and gives

\begin{gather}
    G_{W}(x,y|t) = G_{BD}(x,y|t).
\end{gather}
This observation finishes the proof that the BD state is seen in Static patch as the thermal state. This is exactly the same situation as with Minkwski state in Rindler space.

However, please note that even though the distribution looks like a thermal one, this state cannot be considered as a real thermal state. In fact, in the proper thermal state there is a non-zero Debye mass, but as was shown in \cite{Popov:2017xut}, this mass is actually equal to zero due to the properties of the BD vacua and the de Sitter isometry.

Consider now a generalisation of the above equation to a generic temperature $\beta$:

\begin{gather}
    G_{W\beta}(x,y|t)=2\sum_n\int d\tau \frac{\Im G_{BD}(x,y|\tau)}{t-\tau+i n \beta - i \eps}.
\end{gather}
For the case $\beta_q= \frac{2\pi}{q}$ one can split this expression into two sums

\begin{gather}
    G_{W\beta_q}(x,y|t)=2\sum_n \sum_{m=0}^{q-1}\int d\tau \frac{\Im G_{BD}(x,y|\tau)}{t-\tau+ 2\pi n i+ \frac{2\pi m}{q} i - i \eps} = \sum^{q-1}_{m=0} G_{BD}\left(x,y|\tau+ \frac{2\pi m}{q} i\right).
\end{gather}
Then, there is an interesting case of $q=2,\beta=\pi$. In this case the propagator is

\begin{gather}
    G_{W\beta_2}(x,y|t)=\frac{\pi}{\cosh \pi \mu} P_{-\frac12+i\mu}\left(-\frac{\cosh(t_1 - t_2) + \sinh x_1 \sinh x_2}{\cosh x_1 \cosh x_2}\right) +\notag\\+  \frac{\pi}{\cosh \pi \mu} P_{-\frac12+i\mu}\left(-\frac{-\cosh(t_1 - t_2) + \sinh x_1 \sinh x_2}{\cosh x_1 \cosh x_2}\right),
\end{gather}
the second term brings in additional singularities on the horizon. It would be interesting to study the possible consequences or effects of these singularities.

 \section{Concluding remarks}

We give here a heuristic argument to explain why the time ordered propagator does not have an imaginary part at the coincidence limit in odd dimensions\footnote{ETA would like to thank N.Nekrasov for communicating this argument, which he has learned from A.Polyakov. At least that what ETA remembers from the discussion of this issue with NN in 2011.}. 

It is possible to convert the effective action considered in the Introduction into a quantum mechanical path integral:
\begin{align*}
iS_{eff}=\log\left(\int d[\varphi]e^{i\int d^dx \mathcal{L}}\right)=\int_0^{\infty}\frac{dT}{T}\int_{x(0)=x(T)} d[x] e^{i\int_0^Tdt\left(\frac{\dot{x}^2 }{4}+{m^2}\right)}.
\end{align*}
To evaluate the last integral we may invoke the semiclassical approximation and obtain that
\begin{align*}
iS_{eff}=\int_0^{\infty}\frac{dT}{T} e^{iS_\text{ext}}\sqrt{\frac{1}{\text{det}\left(\Delta\right)}},
\end{align*}
where $S_{\text{ext}}$ is the extremal action for the particle in the background under consideration and $\Delta$ is the operator describing fluctuations around the extremum. 

Usually one calculates the above  integral by a Wick rotation to the Euclidean signature. The Euclidean manifold of the complex de Sitter spacetime is a sphere and the geodesics are the maximal circles. This provides the required  $S_{\text{ext}}$ which is an extremum rather than a minimum. In fact, on the $D$--dimensional sphere there are $(D-1)$ directions along which maximal circles can shrink. Thus, there are $(D-1)$ negative eigenvalues and the effective action contains the contribution
\begin{align*}
{\text{det}\left(\Delta\right)}^{-\frac{1}{2}}\sim (-1)^{\frac{D-1}{2}}.
\end{align*}
Consequently in even dimension Im $(S_{eff}) \neq 0$; on the contrary, in odd dimension there is an even number of negative eigenvalues which results in a real value for the effective action.

In all, there is no imaginary contribution to the effective action in odd--dimensional de Sitter space, but this does not mean that there is no particle production in odd--dimensional de Sitter space \cite{Akhmedov:2013vka}; the Hamiltonian is time dependent and cannot be diagonalized once and forever. As a consequence, in general the notion of particle can be missing, or at best becomes ambiguous.

\textcolor{black}{The main point of this paper is however that the In-Out formalism provides only a hint that there is something interesting going on.    This is well-known from the prototypical case of a time-dependent electric field in flat space, where the particle number at intermediate times generally depends on the choice of a basis of reference states associated with a particular truncation of the adabiatic expansion \cite{dabrowskidunne} (see also \cite{Akhmedov:2014hfa}, \cite{Akhmedov:2014doa}).}
Moreover, as is explained in \cite{Akhmedov:2008pu} the infrared loop divergences do not cancel out in the background fields, unlike the usual case in flat space field theory in Feynman technique. In  non--stationary quantum field theory to calculate loop integrals one has to apply the Schwinger--Keldysh formalism rather than the Feynman In--Out technique  (see \cite{Akhmedov:2017ooy} for a review)  and  calculate stress--energy fluxes rather than transition amplitudes. In such a situation one has to set up an initial state and consider its destiny under the full interacting Hamiltonian evolution. Although the notion of particle is missing, still the flux may be non--trivial. In fact, it happens that loop corrections to the correlation functions are growing with time \cite{Krotov:2010ma}, \cite{Akhmedov:2017ooy}, \cite{Akhmedov:2019cfd}, \cite{Akhmedov:2013vka} and are not suppressed in comparison with tree--level contributions to the stress--energy flux. Moreover, in some situations loop corrections violate the de Sitter isometry and generate non--trivial fluxes \cite{Akhmedov:2019cfd}, \cite{Akhmedov:2013vka}.

We would like to thank A.Polyakov, E.Mottola and N.Nekrasov for useful discussion.
AET, BKV, DDV, UM and CS would like to thank Hermann Nicolai and Stefan Theisen for the hospitality at the Albert Einstein Institute, Golm, where the work on this project was started.
AET thanks the INFN, Sez di Milano and the University of Insubria at Como for hospitality and financial support.
This work was supported by the Russian Ministry of Science and Education, project number 3.9911.2017/BasePart. The work of ETA was supported by the grant from the Foundation for the Advancement of Theoretical Physics and Mathematics “BASIS” and by RFBR grant 18-01-00460 А. The work of FKP was supported by the US NSF under Grant No. PHY-1620059.

\appendix
\section{Appendix}

\label{appendix}

This calculation is applicable only for even dimensions. Here and below if we write a product of two Legendre functions (for example:$\mathsf{P}(x) \mathsf{Q}(y)$) we assume that this product is ordered: $x>y$. To simplify equations below we use such notations as in $\eqref{FS}$. We need the following three equations which were used e.g. in \cite{Fukuma:2013mx}:

\begin{multline}
\label{A1}
\tag{A1}
\frac{\Big(\cosh t_2 \cosh t_1 \Big)^{-\frac{D-1}{2}}}{(D-2)|\Omega_{D-1}|} \sum_j \Big(2j+D-2 \Big) \mathsf{P}^{ -i \mu}_{j+\frac{D-3}{2}} (\tilde{t}_2) \mathsf{Q}^{ i \mu}_{j+\frac{D-3}{2}} (\tilde{t}_1) C_j^\frac{D-2}{2}(\vec{x}\vec{y})=\\=
\frac{(-1)^\frac{D-2}{2}}{2(2\pi)^\frac{D}{2}} \Big[(Z_+^2-1)^{-\frac{D-2}{4}} Q_{i \mu-\frac{1}{2}}^\frac{D-2}{2}(-Z_+)+(Z_-^2-1)^{-\frac{D-2}{4}} Q_{i \mu-\frac{1}{2}}^\frac{D-2}{2}(-Z_-)\Big]=\\=
\frac{(-1)^\frac{D-2}{2}}{2(2\pi)^\frac{D}{2}} \Big[ S_{i\mu-\frac{1}{2}}^\frac{D-2}{2}(Z_+)+S_{i\mu-\frac{1}{2}}^\frac{D-2}{2}(Z_-)\Big],
\end{multline}
\begin{multline}
\tag{A2}
\label{A2}
\frac{\Big(\cosh t_2 \cosh t_1 \Big)^{-\frac{D-1}{2}}}{(D-2)|\Omega_{D-1}|} \sum_j \Big(2j+D-2 \Big)\frac{\Gamma(j+\frac{D-3}{2}+i\mu+1)}{\Gamma(j+\frac{D-3}{2}-i\mu+1)} \mathsf{P}^{-i \mu}_{j+\frac{D-3}{2}} (\tilde{t}_2) \mathsf{Q}^{ -i \mu}_{j+\frac{D-3}{2}} (\tilde{t}_1) C_j^\frac{D-2}{2}(\vec{x}\vec{y})=\\=
\frac{(-1)^\frac{D-2}{2}}{2(2\pi)^\frac{D}{2}} \Big[e^{\pi\mu}(Z_+^2-1)^{-\frac{D-2}{4}} Q_{i \mu-\frac{1}{2}}^\frac{D-2}{2}(-Z_+)+e^{-\pi\mu}(Z_-^2-1)^{-\frac{D-2}{4}} Q_{i \mu-\frac{1}{2}}^\frac{D-2}{2}(-Z_-)\Big]=\\=
\frac{(-1)^\frac{D-2}{2}}{2(2\pi)^\frac{D}{2}} \Big[ e^{\pi\mu}S_{i\mu-\frac{1}{2}}^\frac{D-2}{2}(Z_+)+ e^{-\pi\mu}S_{i\mu-\frac{1}{2}}^\frac{D-2}{2}(Z_-)\Big]
\end{multline}
and
\begin{multline}
\tag{A3}
\label{A3}
\frac{\Big(\cosh t_2 \cosh t_1 \Big)^{-\frac{D-1}{2}}}{(D-2)|\Omega_{D-1}|} \sum_j \Big(2j+D-2 \Big) \mathsf{Q}^{ -i \mu}_{j+\frac{D-3}{2}} (\tilde{t}_2) \mathsf{Q}^{ i \mu}_{j+\frac{D-3}{2}} (\tilde{t}_1) C_j^\frac{D-2}{2}(\vec{x}\vec{y})=\\=
-\frac{i\pi}{2}\frac{(-1)^\frac{D-2}{2}}{2(2\pi)^\frac{D}{2}} \bigg[(Z_+^2-1)^{-\frac{D-2}{4}} Q_{i \mu-\frac{1}{2}}^\frac{D-2}{2}(-Z_+)-(Z_-^2-1)^{-\frac{D-2}{4}} Q_{i \mu-\frac{1}{2}}^\frac{D-2}{2}(-Z_-)-\\-
\frac{\pi}{\cosh \pi\mu }\Big((Z_+^2-1)^{-\frac{D-2}{4}} P_{i \mu-\frac{1}{2}}^\frac{D-2}{2}(Z_+)-(Z_-^2-1)^{-\frac{D-2}{4}} P_{i \mu-\frac{1}{2}}^\frac{D-2}{2}(Z_-)\Big)\bigg]=\\=
-\frac{i\pi}{2}\frac{(-1)^\frac{D-2}{2}}{2(2\pi)^\frac{D}{2}} \bigg[S_{i\mu-\frac{1}{2}}^\frac{D-2}{2}(Z_+)-S_{i\mu-\frac{1}{2}}^\frac{D-2}{2}(Z_-)-
\frac{\pi}{\cosh \pi\mu }\Big(F_{i\mu-\frac{1}{2}}^\frac{D-2}{2}(Z_+)-F_{i\mu-\frac{1}{2}}^\frac{D-2}{2}(Z_-)\Big)\bigg].
\end{multline}
Where we use $Z_\pm$, which is defined in $\eqref{ZZ}$. To compute $(\ref{propalphabeta})$ we need to perform four summations over $j$.  According to \cite{Fukuma:2013mx}:
\begin{align}
\tag{A4}
\label{A01}
\mathsf{P}_\nu^{i \mu}(x)=\frac{\Gamma(\nu+i\mu+1)}{\Gamma(\nu-i\mu+1)}\Big[\cosh \pi \mu \  \mathsf{P}_\nu^{-i\mu}(x)+\frac{2i}{\pi} \sinh\pi \mu \  \mathsf{Q}_\nu^{-i\mu} \big],
\end{align}
\begin{align}
\label{A02}
\tag{A5}
\mathsf{Q}_\nu^{-i \mu}(x)=\frac{\Gamma(\nu+i\mu+1)}{\Gamma(\nu-i\mu+1)}\Big[ -\frac{\pi i}{2} \sinh\pi \mu \ \mathsf{P}_\nu^{-i\mu}(x)+ \cosh \pi \mu \  \mathsf{Q}_\nu^{-i\mu} \big].
\end{align}
One can transform sums in $(\ref{propalphabeta})$ into $(\ref{A1}), \ (\ref{A2}), \  \text{and} \ (\ref{A3})$. We denote sums as follows:
\begin{align*}
\mathsf{P}^+\mathsf{Q}^- = \frac{\Big(\cosh t_2 \cosh t_1 \Big)^{-\frac{D-1}{2}}}{(D-2)|\Omega_{D-1}|} \sum_j \Big(2j+D-2 \Big) \mathsf{P}^{ -i \mu}_{j+\frac{D-3}{2}} (\tilde{t}_2) \mathsf{Q}^{ i \mu}_{j+\frac{D-3}{2}} (\tilde{t}_1) C_j^\frac{D-2}{2}(\vec{x}\vec{y}).
\end{align*}
And the results of transformations is as follows:
\begin{align*}
\mathsf{P}^+ \mathsf{P}^-=-\frac{2}{\pi}\frac{i}{\sinh\mu\pi} (\ref{A2})+\frac{2 i \cosh\mu\pi}{\pi \sinh\mu\pi}  (\ref{A1}),
\end{align*}
and
\begin{align*}
\mathsf{Q}^+ \mathsf{P}^-=-\frac{1}{\sinh^2 \mu\pi} (\ref{A1})^*+\frac{\cosh \mu\pi}{\sinh^2 \mu\pi} (\ref{A2})+\frac{2 i \cosh \mu\pi}{\pi \sinh \mu \pi} (\ref{A3}).
\end{align*}

Where $(A1)$, $(A2)$ and $(A3)$ means the expressions from the corresponding equations.

\end{document}